\documentclass[12pt]{article}

\usepackage[english]{babel}
\usepackage[T1]{fontenc}

\usepackage[margin=1in]{geometry}

\usepackage{amssymb,amsmath,amsthm}
\usepackage{tikz}
\usepackage{graphicx,ctable,booktabs}
\usepackage[colorinlistoftodos]{todonotes}
\usepackage{bbm}
\usepackage{enumerate}
\usepackage{mathrsfs}
\usepackage{pdflscape}
\usepackage{tabularx}
\usepackage{subfigure}
\usepackage{natbib}
\usepackage{placeins}
\usepackage[hang,flushmargin]{footmisc}
\usepackage[capposition=top]{floatrow}
\usepackage{setspace}
\usepackage{color}
\usepackage{lscape}
\usepackage{comment}
\usepackage{dsfont}
\usepackage{verbatim}
\usepackage{multirow}
\usepackage{longtable}
\usepackage{booktabs}
\usepackage{natbib}

\usepackage{amsmath}
\usepackage{amsfonts}
\usepackage{amssymb}
\usepackage{tikz}
\usetikzlibrary{decorations.pathreplacing}
\usepackage{physics}
\usepackage{graphicx}%
\usepackage[utf8]{inputenc}
\usepackage{varioref}
\usepackage[hidelinks]{hyperref}
\usepackage{enumitem}
\setlist[enumerate]{wide=0pt, widest=99,leftmargin=\parindent, labelsep=* } 
\usepackage{indentfirst}
\usepackage{threeparttable}
\usepackage{graphics}
\usepackage{verbatim} 
\usepackage{threeparttable}
\usepackage{rotating}
\usepackage[capposition=top]{floatrow}
\usepackage{caption}
\usepackage{mathtools}
\usepackage{bbm}
\usepackage[ruled,vlined]{algorithm2e}

\usepackage{xcolor,soul} 
\usepackage{diagbox} 
\usepackage{rotating}
\usepackage{adjustbox}
\usepackage{url}

\newtheorem{theorem}{Theorem}
\newtheorem{proposition}[theorem]{Proposition}

\hypersetup{
   colorlinks=true, 
   linktoc=all,        
   linkcolor=black,  
   urlcolor  = blue,
   citecolor = black,
}

\begin{document}
\begin{titlepage}
\onehalfspacing
\title{%
\Large A General Equilibrium Study of Venture Capitalists' Effort on Entrepreneurship
}

\author{{Liukun Wu} 
}
\date{September 2024}
\maketitle
\thispagestyle{empty}
\bigskip

\centerline{\bf Abstract}

\begin{singlespace}
In this paper, I propose a new general equilibrium model that explains stylized facts about venture capitalists' impact on their portfolio firms. Venture capitalists can help increase firms' productivity, yet they face increasing entry costs to enter. I characterize steady state effort choice, entry threshold, and mass of venture capitalists, and show how they are affected by change in upfront investment, interest rate, and entry costs. The key contribution is that public policy to stimulate startups by subsidizing upfront investments or reducing interest cost have limited success if not accompanied by an increasing supply of experts who can improve business ideas.

\end{singlespace}

\end{titlepage}

\section{Introduction}
Venture capitalists play a prominent role in the US economy, tracing back to as early as the 19th Century. The earliest venture capitalists invested in the whaling industry, which at that time had high returns on capital but was both risky and complex in organizational structure. Nowadays, venture capitalists' influence has already extended to quite a few industries. Venture capitalists play an active role in innovative startups in information technology, medical sciences, communications, and so on. They first screen and evaluate potential startups for their future value. Then, they sign a contract with founders that stipulates the amount of funding and allocation of shares. After funding has been injected, venture capitalists go through several rounds to further evaluate startups, provide managerial expertise to improve business operations, and provide further funding, or withdraw investments, depending on the status of startups. Startups that eventually succeed will go through an exit event of either an IPO or a merger and acquisition, and venture capitalists will cash out their shares at that time.

Given the amount of uncertainty and information asymmetry in the early stage of startups, how to write efficient contracts (initially and subsequently) is often the focus of venture capitalist literature. Contracts that allocates funds to "good" projects and incentivizes entrepreneur's effort will have a positive effect on aggregate total factor productivity (TFP), as resources are allocated more efficiently (\cite{greenwood_financing_2018}). What is often neglected, however, is that venture capitalists' managerial expertise also enhances startups' TFP, by making business operations more efficient. This is largely due to the lack of data. Venture capitalists are very discreet about their operations. Moreover, their actions are often state-contingent. Thus, it is difficult to assemble a dataset that records what actions have been taken on startups and their outcomes. 

This paper leverages a key stylized fact obtained from the World Management Survey, a rare dataset that gathers management performance of both venture and non-venture-funded firms, which shows that venture-funded firms have a higher management score, specifically in setting business targets and hiring policies. This allows me to motivate a production function in which venture capitalists' effort enhances firms' TFP and study its macroeconomic impact.

The key contribution of this paper is that it is the first general equilibrium model about venture capitalists' human capital and its supply. The key assumption is that venture capitalists allocate their effort to firms with heterogeneous productivities under a supply constraint, and their effort enhances firms' TFP. 

The results of the model align with empirical observations. It correctly predicts the selection of high-productivity startups, and higher employment levels of venture-funded firms in the long run. Moreover, the comparative statics are also in line with empirical findings in literature. For example, a drop in initial investment increases the number of marginal projects funded, induces more venture capitalists to enter, and increases aggregate labor productivity. Therefore, this model can be used as a benchmark theoretical model about venture capitalists' effort allocation.

The limitation of this study is that it is unable to quantify the increase in aggregate TFP brought by venture capitalists. In order to do that, a micro-level dataset on venture capitalists' time use is needed. Moreover, moments on the annual number of employees in the venture capital industry are also essential to determine the entry dynamics. It is thus left for future study. 

\section{Literature Review}
This paper is related to three strands of literature. First is the theory for determining the contract between venture capitalists and entrepreneurs. Earlier papers focus on the allocation of control rights. The contract mainly resolves the conflicts of future actions on the firm (\cite{berglof_control_1994}), or hold-up problem (\cite{hellmann_allocation_1998}). Papers later on start to understand information frictions between venture capitalists and entrepreneurs. \cite{bergemann_venture_1998} discusses how to determine contracts when venture capitalists learn about project's type over time. This paper is closest to the strand of literature that focuses on double moral hazard (\cite{casamatta_financing_2003}, \cite{repullo_venture_2004}). Similar to other papers' setup, my model features unobservable effort by both venture capitalists and entrepreneurs. The difference is that venture capitalists' effort increases the productivity of the project, yet there is a supply limit. As a result, the optimal contract between venture capitalists and entrepreneurs is determined by the optimal allocation of effort among projects with different productivities.

This paper also touches upon the difference between banks and venture capitalists. In \cite{ueda_banks_2004}, venture capitalists have a better screening technology, so that they can fund entrepreneurs with little collateral or projects that require large upfront investments. In \cite{winton_entrepreneurial_2008}, venture capitalists are better at assessing uncertainty in firms' business strategies, and thus are suited for funding riskier projects with skewed payoffs. In this paper, venture capitalists play a role that goes beyond passive assessment. Instead, they possess managerial expertise that can even enhance entrepreneur's own effort to build up business organizations. Their effort will positively affect output even in the long run.  

This paper is also the first general equilibrium model that endogenizes the supply of venture capitalists' effort by modeling the entry dynamics of both entrepreneurs and venture capitalists. My model abstracts from search frictions, discussed in \cite{silveira_venture_2016}, in order to focus on how the supply of venture capitalists' effort affects macroeconomic quantities. \cite{opp_venture_2019} focuses on the cyclicality of venture capitalists' investments and how it affects growth. My model mainly focuses on steady state comparisons. \cite{greenwood_financing_2018} incorporates dynamic contracts into an endogenous growth framework with both asymmetric information and moral hazard problems, whereas my model chooses to focus on moral hazard. \cite{akcigit_synergizing_2019} also talks about the complementality between venture capitalists and entrepreneurs, but does not consider the variable intensity of venture capitalists' effort. 

\section{Stylized Facts}
This section provides a picture of venture-funded firms in the macro-economy, capturing key facts that the general equilibrium model predicts. I seek to understand the characteristics of venture-funded firms upon selection, and how venture-funded firms compare to those non-venture-funded ones. Despite data limitations, I try to provide some evidence of the supply of venture capitalists' effort.

I combine two datasets for the analysis. The dataset that allows me to track all firms over time is the National Establishment Time Series (NETS) based on Dun \& Bradstreet microdata\footnote{D\&B has strong profit incentives to compile accurate data, as they use and sell the data for marketing and credit scoring. They collect data from state secretaries, Yellow Pages, court records, credit inquiries, and direct telephone contact.}. It provides detailed annual establishment-level information on identifiers, industry, employment, sales, and so on.\footnote{Although the US Census Bureau's Longitudinal Business Database (LBD) is a better source of data, it is not accessible during the pandemic period. A comparison between LBD and NETS is discussed in Appendix \ref{ch2_appendix}.} Firms that receive venture funding are identified using the VentureXpert dataset, which contains round-level information on the funded firm, funding amount, and the identity of venture capitalists. Firms in VentureXpert are matched with NETS using name and other identifier matching techniques, and details can be found in Appendix \ref{ch2_appendix}.

\subsection{Characteristics of Venture-Funded Firms}
\begin{figure}[H]
	\begin{center}
		\includegraphics[width=0.8\textwidth]{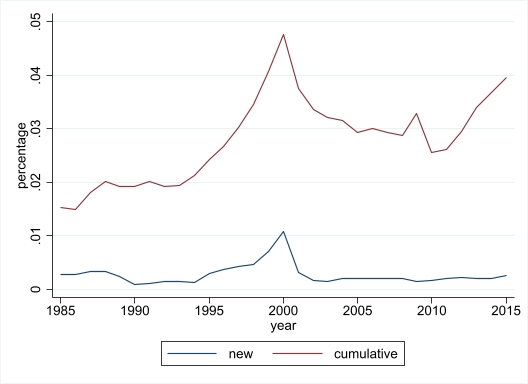}
		\caption{Percentage of Venture-Funded Firms in the Economy}
		\label{ch2_fig11}
	\end{center}
\end{figure}

The number of firms funded by venture capitalists is very small and concentrated in a few industries. Figure \ref{ch2_fig11} shows the percentage of venture-funded firms (or establishments) found in NETS. Less than 0.01\% of the firms are newly funded by venture capitalists every year. The total number of firms funded by venture capitalists at some point in time also does not exceed 0.05\%. In VentureXpert, 65\% of the firms are in the information technology sector. Among the rest, medical and life science firms comprise another 23\%.

\begin{figure}[H]
	\begin{center}
		\includegraphics[width=0.8\textwidth]{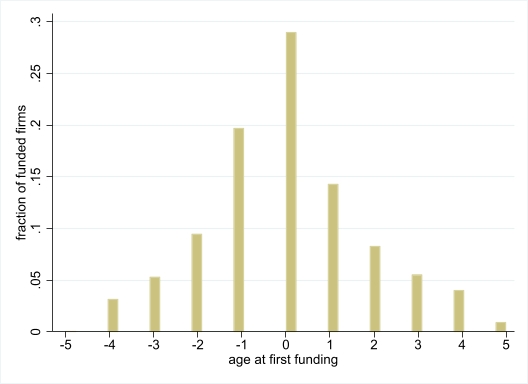}
		\caption{Age Distribution of Venture-Funded Firms}
		\label{ch2_fig12}
	\end{center}
\end{figure}

Firms funded by venture capitalists are also very young. In Figure \ref{ch2_fig12}, most firms are at Age 0 when they first receive venture funding, where Age 0 is the year that an establishment is first observed in NETS. Some firms are funded even earlier before they get recorded in NETS as an employer. 

\begin{figure}[H]
	\begin{center}
		\includegraphics[width=0.8\textwidth]{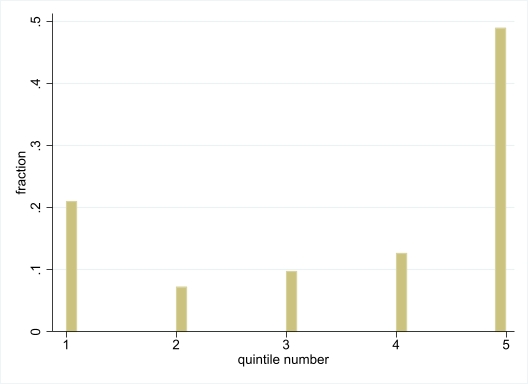}
		\caption{Employment Quintile of Venture-Funded Firms}
		\label{ch2_fig13}
	\end{center}
\end{figure}

At the first year of their funding, venture-funded firms are much larger in employment compared to their peers. Figure \ref{ch2_fig13} shows that almost 50\% of the venture-funded firms are in the highest quintile when they were selected.\footnote{There is too much missing data to compute employment growth quintile for venture-funded firms.} This suggests that venture capitalists are highly selective in their investments, preferring to work with firms that exhibit high employment in their early years.

\subsection{Comparison of Venture-Funded and Non-Venture-Funded Firms}
\begin{figure}[H]
	\begin{center}
		\includegraphics[width=0.8\textwidth]{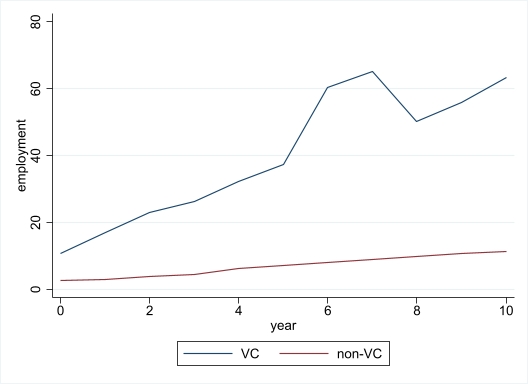}
		\caption{Employment After Venture Funding}
		\label{ch2_fig14}
	\end{center}
\end{figure}

Now I examine the effects of venture capitalist involvement post-funding. Figure \ref{ch2_fig14} plots the evolution of average employment after the first venture funding year for both venture- and non-venture-funded firms, where year 0 is the first year that a firm receives venture funding. The two groups both exhibit an upward trend in employment. However, venture-funded firms grow to a much larger size following first funding. At the end of year 5, a venture-funded firm is almost 4 times larger than a non-venture-funded firm on average. 

To control for observables, I run the regression of post-funding employment on venture-funded status, controlling for the initial size of the firm, state, and industry. Table \ref{ch2_table1} shows the results for employment at 3, 5, and 10 year horizon. Firms funded by venture capitalists on average will be 70.5\% larger after 3 years of funding, 87.8\% larger after 5 years, and almost twice the size after 10 years. This suggests that venture capitalist involvement in firms is likely to have a strong effect. 

\begin{figure}
	\begin{center}
		\includegraphics[width=0.8\textwidth]{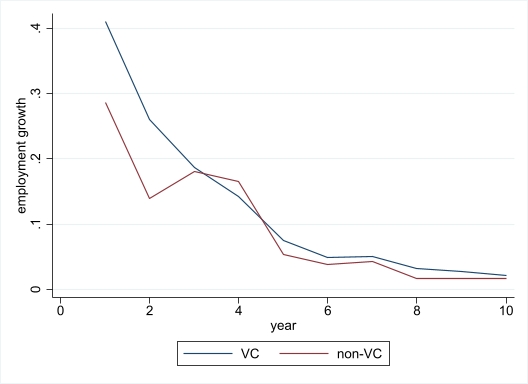}
		\caption{Employment Growth After Venture Funding}
		\label{ch2_fig15}
	\end{center}
\end{figure}

I also plot the evolution of employment growth for venture- and non-venture-funded firms. 30\% of observations is missing employment growth, so there is significant noise. Figure \ref{ch2_fig15} shows that venture-funded firms have higher employment growth after funding.\footnote{\cite{akcigit_synergizing_2019} documents significantly higher employment growth of venture-funded firms using Census data, which has better data quality especially for younger-aged establishments.} 
\begin{table}[H]
	\centering
	\def\sym#1{\ifmmode^{#1}\else\(^{#1}\)\fi}
	\caption{Regression of Long Term Employment on Venture Status}
	\label{ch2_table1}
	\begin{tabular}{lccc}
		
		\hline \hline 
		&Emp(3yr,ln)&Emp(5yr,ln)&Emp(10yr,ln) \\
		\hline
		VC(Dummy)            &       0.705\sym{***}&       0.878\sym{***}&       1.054\sym{***}\\
		&    (0.025)         &    (0.033)         &    (0.035)         \\
		Employment at first year (ln)              &       0.773\sym{***}&       0.713\sym{***}&       0.676\sym{***}\\
		&   (0.008)         &   (0.008)         &   (0.008)         \\
		Constant            &       0.295\sym{***}&       0.387\sym{***}&       0.462\sym{***}\\
		&   (0.006)         &   (0.006)         &   (0.008)         \\            
		\hline
		State F.E. & Yes & Yes & Yes \\
		Industry F.E. & Yes & Yes & Yes \\
		Year F.E. & Yes & Yes & Yes \\
		$N$        &    24,213,551         &    16,486,433         &     7,474,253         \\
		$R^2$           &       0.619         &       0.554         &       0.515         \\
		\hline\hline
	\end{tabular}
	\begin{tablenotes}
		\item \footnotesize Standard errors in parentheses, \sym{*} \(p<0.05\), \sym{**} \(p<0.01\), \sym{***} \(p<0.001\)
	\end{tablenotes}
\end{table}

\subsection{Supply of Venture Capitalists' Effort}
Since venture capitalists' involvement heavily influence firms' performance, it is important to understand the total supply of venture capitalists' effort every year, and how venture capitalists allocate their time to different firms. Ideally, we would like to have a dataset that tracks the number of investment staff at each venture capital firm every year\footnote{The Preqin Private Equity dataset has a section with the number of investment staff at each fund, However, it is only the most up-to-date staff number count and does not vary by year.}. We would also like to have a time use dataset that records the number of hours spent on each firm. Such a dataset, however, is difficult to obtain. Therefore, to understand the supply of venture capitalists' effort, I will use venture capital firm (which can include multiple funds and fund managers) as the unit of measure and examine its number over the years.

\begin{figure}
	\begin{center}
		\includegraphics[width=0.8\textwidth]{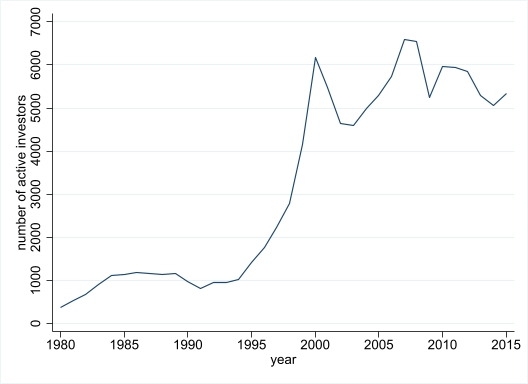}
		\caption{Number of Active Investors by Year}
		\label{ch2_fig16}
	\end{center}
\end{figure}

Figure \ref{ch2_fig16} shows the number of active investors from 1980 to 2015. An active investor is a venture capital firm that has made at least one investment in that year. The number of active investors is relatively constant at about 1000 before the Internet boom, and hovers around 5500 after the boom. There are two possible reasons that equilibrium supply of investors varies little in the short term. On the demand side, it is possible that the number of good firms is limited, until a large technological shock happens. On the supply side, venture capital funds may face capacity constraints if fund manager's human capital is not easily scalable and it takes time to train a new, qualified fund manager.

Some anecdotal evidence leans towards a supply capacity constraint interpretation. In \cite{bernstein_impact_2016}, the authors note that, "investors are time constrained and notoriously reluctant to provide data on their operations". Moreover, their paper shows that the introduction of new airline routes that reduce venture capitalists' travel times to their existing portfolio companies increases onsite involvement. This suggests that venture capitalists might be facing a capacity constraint for the amount of time available to engage with portfolio companies. Once that constraint is loosened, venture capitalists are able to put in more effort.

In terms of work hours, \cite{gompers_venture_2020} provides a glimpse into the amount of time allocated to different types of firms and different activities. In terms of time use, venture capitalists work on average 55 hours a week. The majority of hours is split between sourcing deals (15 hours) and assisting portfolio companies (18 hours). Among the 900 venture capital firms being surveyed, the average number of hours spent on information technology and healthcare companies is similar, at about 20 hours. Thus, I deduce that venture capitalists face a capacity constraint, and the optimal amount of effort allocated to companies is less likely to depend on industry type. 

\subsection{Summary}
Below are the key observations that my general equilibrium model seeks to capture.
\begin{enumerate}
	\item Venture capitalists selectively invest in young firms with high employment.
	\item Venture capitalists' involvement increases firms' employment even in the long run.
	\item Following a positive technological shock, the equilibrium supply of venture capitalists increases.  
\end{enumerate}

\section{Partial Equilibrium}
I first illustrate fundamental forces that influence venture capitalists' effort provision and financing decision using a two period model in partial equilibrium. The key element of the model is the production function that describes how venture capitalists' effort increases total factor productivity. 

\subsection{Model Environment}
Consider an economy with two dates ($t=0,1$), a large set of penniless risk-neutral entrepreneurs, a group of risk-neutral venture capitalists, and a risk-neutral bank. Entrepreneurs have innovative projects that require an upfront investment $I$, which can come from the bank or venture capitalists. Both venture capitalists and the bank can raise funds at risk free interest rate $r$.

Entrepreneurs are heterogeneous with observable types $(z,c)$ distributed as $\Phi(z,c)$. The production function is given by,

\begin{equation}
	\label{eqn_production}
	y = z^\chi F^\theta l^\beta, \qquad F = [\gamma h^{\frac{\sigma-1}{\sigma}}
	+ f^{\frac{\sigma-1}{\sigma}} ]^{\frac{\sigma}{\sigma-1}}, \chi + \theta + \beta = 1
\end{equation}
where $z$ represents total factor productivity (excluding management), $F$ represents management productivity, and $l$ is labor, which is supplied at wage $w$. Management can come from either entrepreneurs, using $f$ units of effort at cost $c$, or from venture capitalists, using $h$ units of effort. The importance of venture capitalists' effort is $\gamma$, and the degree of substitutability between venture capitalists' effort and entrepreneurs' effort is $\sigma$. I assume here that $\sigma > 1$, i.e. $h$ and $f$ are gross substitutes.\footnote{Literature notes substantial management turnover at venture-funded firms (\cite{hellmann_venture_2002},\cite{kaplan_should_2009-1}). Either venture capitalists or new CEOs will be able to replace founders to continue business operations.} The probability of success for all projects is constant at $\epsilon$.

Each entrepreneur is matched with a particular venture capitalist, and venture capitalist can work with many entrepreneurs. There is no search friction. Upon the match, venture capitalists observe $(z,c)$ and provide $h$ units of effort. The total supply of effort is $H$. The contract between venture capitalists and entrepreneurs is determined by Nash Bargaining. It specifies a payment $p$ to venture capitalists if the project is successful, with venture capitalists' bargaining power being $\alpha$. Effort levels are not observable and therefore not contractible.

The sequence of events is as follows:
\begin{itemize}
	\item At $t=1$, entrepreneurs draw $(z,c)$ and decide whether or not to undertake the project, and where to seek financing. If entrepreneur goes to the bank, he receives investment $I$, and then puts effort $f$ to work on the project on his own. If entrepreneur goes to the venture capitalist, the two of them first sign a contract specifying $p$, and $I$ is invested. Then, entrepreneur and venture capitalist each chooses his effort level.
	\item At $t=2$, project outcome is realized. Successful project produces output according to \ref{eqn_production}, whereas unsuccessful project has zero remaining value. Entrepreneurs repay the bank and venture capitalists.
\end{itemize}

\subsection{Optimal Contracts and Financing Decision}
Given wage $w$, labor choice $l$ of an ex post successful project is given by the following profit maximization problem,
\begin{equation}
	\Pi \equiv \max_l \{z^\chi F^\theta l^\beta - wl \}
\end{equation}
The first order condition with respect to $l$ gives,
\begin{equation}
	\label{eqn_labor}
	l = \Big (\frac{\beta}{w}\Big )^{\frac{1}{1-\beta}} z^{\frac{\chi}{1-\beta}} F^{\frac{\theta}{1-\beta}}
\end{equation}
so that profit is
\begin{equation}
	\Pi = (1-\beta)\Big (\frac{\beta}{w}\Big )^{\frac{\beta}{1-\beta}}z^{\frac{\chi}{1-\beta}} F^{\frac{\theta}{1-\beta}}
\end{equation}
\subsubsection*{Bank Finance}
Faced by a zero-profit constraint, the bank will charge an interest rate $i$ so that it receives an expected repayment of $I$. When the project fails, entrepreneurs default on the principal and interest and repay zero. Thus, the interest rate $i$ satisfies,
\begin{equation}
	1 + i = \frac{1+r}{\epsilon}
\end{equation}
Given bank's interest rate $i$, bank-financed entrepreneurs solve the following problem to maximize the present value of profit at $t=0$,
\begin{equation}
	\max_f  \frac{\epsilon}{1+r}(1-\beta)\Big (\frac{\beta}{w}\Big )^{\frac{\beta}{1-\beta}}z^{\frac{\chi}{1-\beta}} F^{\frac{\theta}{1-\beta}} - cf - I , \qquad F=f	
\end{equation}
Solving for the first order condition with respect to $f$, entrepreneurs' effort choice is
\begin{equation}
	f = z \Big ( \frac{\epsilon\theta}{c(1+r)}\Big )^{\frac{\theta+\chi}{\chi}} \Big (\frac{\beta}{w}\Big )^{\frac{\beta}{\chi}}
	\label{eqn_fsolo}
\end{equation}
Present value of the project is thus,
\begin{equation}
	\label{eqn_vsolo}
	E \equiv \max \Big \{ z\chi \Big (\frac{\theta}{c} \Big)^{\frac{\theta}{\chi}} \Big (\frac{\epsilon}{1+r} \Big)^{1+\frac{\theta}{\chi}}\Big (\frac{\beta}{w} \Big)^{\frac{\beta}{\chi}} - I, 0 \}
\end{equation}
since entrepreneurs always have the outside option of not starting the project after the draw of $(z,c)$. Entry threshold for a bank-financed entrepreneur is therefore,
\begin{equation}
	\label{eqn_pcsolo}
	z_s = \frac{I}{\chi}\Big (\frac{\theta}{c} \Big)^{-\frac{\theta}{\chi}} \Big (\frac{\epsilon}{1+r} \Big)^{-1-\frac{\theta}{\chi}}\Big (\frac{\beta}{w} \Big)^{-\frac{\beta}{\chi}}
\end{equation}

\subsubsection*{Venture Capital Finance}
The contract between venture capitalists and entrepreneurs is determined by the following Nash Bargaining problem,
\begin{equation}
	\max_{p}[\frac{\epsilon}{1+r}(1-\beta)\Big (\frac{\beta}{w}\Big )^{\frac{\beta}{1-\beta}}z^{\frac{\chi}{1-\beta}} F^{\frac{\theta}{1-\beta}} -\epsilon p-E]^{1-\alpha}[\epsilon p - I]^\alpha, \quad F = [\gamma h^{\frac{\sigma-1}{\sigma}}
	+ f^{\frac{\sigma-1}{\sigma}} ]^{\frac{\sigma}{\sigma-1}}	
\end{equation}
where $E$, is the outside option for the entrepreneur, that he can choose to finance with the bank or not start the project. Venture capitalists' outside option is $I$, in that they can invest the funds in another project.

Solving for the first order condition for $p$, the surplus for entrepreneur and venture capitalists respectively, is,
\begin{align}
	\label{valuee}
	S_v &\equiv \epsilon p - I = \alpha(\frac{\epsilon}{1+r}(1-\beta)\Big (\frac{\beta}{w}\Big )^{\frac{\beta}{1-\beta}}z^{\frac{\chi}{1-\beta}} F^{\frac{\theta}{1-\beta}} - I - E )    \\
	S_e &= (1-\alpha)(\frac{\epsilon}{1+r}(1-\beta)\Big (\frac{\beta}{w}\Big )^{\frac{\beta}{1-\beta}}z^{\frac{\chi}{1-\beta}} F^{\frac{\theta}{1-\beta}} - I - E )  
	\label{valuevc}
\end{align}

After the contract is signed and $I$ is invested, entrepreneurs and venture capitalists choose their effort level to maximize their expected surplus, given the contract and their rational expectation of the effort of the other. The incentive-compatible effort level for the entrepreneur is,
\begin{equation}
	\max_f (1-\alpha)(\frac{\epsilon}{1+r}(1-\beta)\Big (\frac{\beta}{w}\Big )^{\frac{\beta}{1-\beta}}z^{\frac{\chi}{1-\beta}} F^{\frac{\theta}{1-\beta}} - I - E ) - cf, \qquad F = [\gamma h^{\frac{\sigma-1}{\sigma}}
	+ f^{\frac{\sigma-1}{\sigma}} ]^{\frac{\sigma}{\sigma-1}}
\end{equation}

Similarly, venture capitalists' incentive-compatible effort maximizes his surplus, taking into account entrepreneur's effort, and the capacity constraint at $H$.
\begin{align}
	\max_h \alpha(\frac{\epsilon}{1+r}(1-\beta)\Big (\frac{\beta}{w}\Big )^{\frac{\beta}{1-\beta}}z^{\frac{\chi}{1-\beta}} F^{\frac{\theta}{1-\beta}} - I - E ) \\
	\int_{z\geq z_v,c\geq c_v} h(z,c) d\Phi(z,c) = H
\end{align}
where $z_v,c_v$ denote the threshold values for financing with venture capitalists. Let $v$ denote the shadow cost of venture capitalists' effort, solving for the first order conditions with respect to $f$ and $h$, optimal levels of effort are,
\begin{align}
	\label{eqn_fvc}
	f &= \Big (\frac{v}{c}\frac{1-\alpha}{\alpha \gamma}  \Big )^\sigma h \\
	\label{eqn_hvc}
	h &= z\Big ( \frac{\alpha \epsilon\theta}{v(1+r)}\Big )^{\frac{\theta+\chi}{\chi}} \Big (\frac{\beta}{w}\Big )^{\frac{\beta}{\chi}}\gamma^{\frac{\theta \sigma}{\chi(\sigma-1)}}\Big [ 1+\gamma^{-\sigma}\Big(\frac{v}{c}\frac{1-\alpha}{\alpha} \Big)^{\sigma-1}  \Big]^{\frac{\theta}{\chi(\sigma-1)}-1}
\end{align}
Lastly, the participation constraint for entrepreneurs, is
\begin{equation}
	\label{eqn_pcvc}
	\frac{\epsilon}{1+r}(1-\beta)\Big (\frac{\beta}{w}\Big )^{\frac{\beta}{1-\beta}}z^{\frac{\chi}{1-\beta}} F^{\frac{\theta}{1-\beta}} - I \geq E
\end{equation}
So that total surplus should be larger than zero.

\begin{proposition}
	\label{prop1}
	Venture capitalists' level of effort, $h$, is increasing in $z$ and $c$, and increasing in $H$ (decreasing in $v$).
\end{proposition}
From the production function, entrepreneur's total factor productivity, $z$, and management productivity are complements. The better entrepreneur's idea (higher $z$), the higher the marginal product of venture capitalists' effort $h$, and therefore a higher level of $h$ will be required to equalize the shadow cost $v$. Secondly, venture capitalists' effort is a substitute for entrepreneur's effort ($\sigma>1$). A less efficient entrepreneur (higher $c$) will cause effort to be shifted toward venture capitalists. Lastly, a smaller supply of venture capitalists' effort $H$ makes it more costly to seek venture capitalists' effort. Thus, a higher marginal product of $h$ is needed to equalize the higher shadow cost $v$, hence lowering effort level $h$.

\begin{figure}[H]
	\begin{center}
		\includegraphics[width=0.8\textwidth]{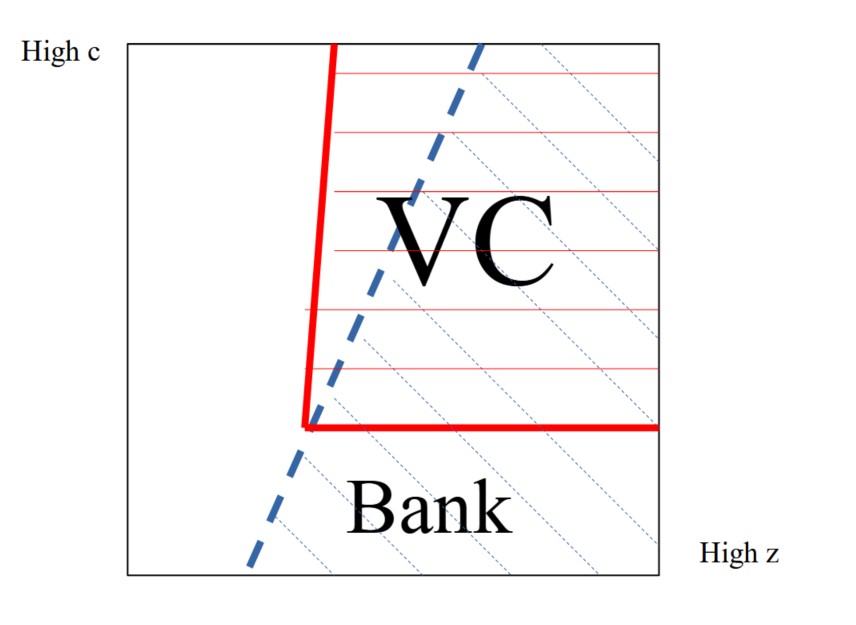}
		\caption{Feasibility Region for VC and Bank Financing}
		\label{ch2_fig1}
	\end{center}
\end{figure}

Figure \ref{ch2_fig1} depicts the feasibility region for bank and venture capitalist financing respectively. More details can be found in Appendix \ref{ch2_appendix}. The area in the blue dashed region is the range of projects that will be profitable for entrepreneurs if funded by banks. For an entrepreneur with a higher cost of effort $c$, his project has a lower management productivity, and therefore requires a higher $z$ to produce enough output to repay the upfront investment. The area in the red solid region is the range of projects that are profitable for venture capitalist funding. It consists of two parts. First, projects that are feasible under bank financing are now more profitable when financed by venture capitalists. Second, some projects that are not feasible under bank financing are now profitable with venture capitalists. This corresponds to the red-shaded area in between the red solid line and the blue dashed line. This is because venture capitalists' effort is more efficient (lower shadow cost $v$) than some entrepreneurs' effort cost $c$. Working with venture capitalists yields a higher management productivity than working as a solo entrepreneur, and the output is higher.

\subsection{Empirical Implications}
Results from the partial equilibrium model have the following empirical implications.

\begin{figure}[H]
	\begin{center}
		\includegraphics[width=0.8\textwidth]{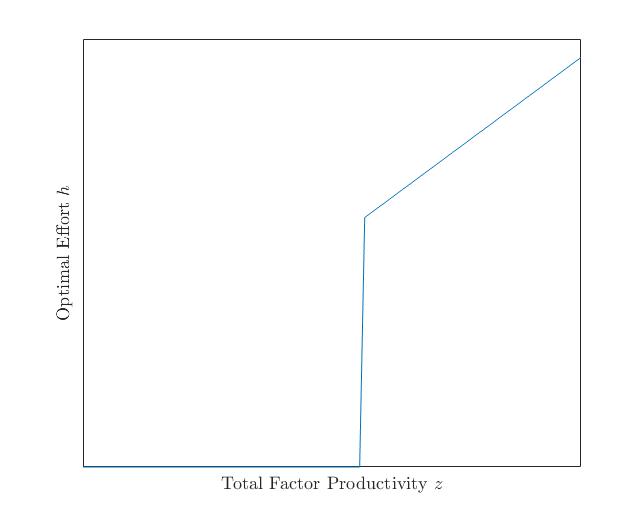}
		\caption{Optimal Effort Choice $h$ by Total Factor Productivity $z$}
		\label{ch2_fig2}
	\end{center}
\end{figure}

\textit{A. Highly innovative startups get funded by venture capitalists.}\\
Figure \ref{ch2_fig2} shows the model predictions on effort choice varying by total factor productivity, for a given founder's effort cost. Only projects with high enough productivity are funded by venture capitalists, and more productive projects receive more effort. It is commonly observed that venture capitalists concentrate their funding in highly innovative startups, mostly from the information technology and biotechnology sectors. Moreover, startups that turn out to be exceptionally successful, such as Square, Uber, and Snapchat, often have board members from quite a few venture capital funds. Anecdotes from professionals suggest that venture capitalists tend to devote more time to larger startups.

\begin{figure}[H]
	\begin{center}
		\includegraphics[width=0.8\textwidth]{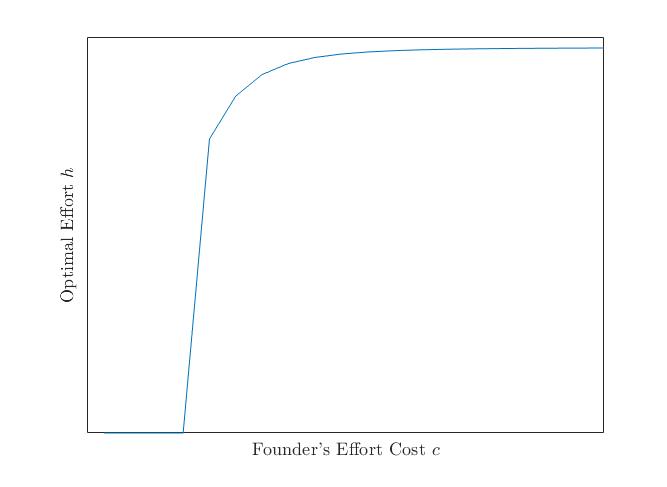}
		\caption{Optimal Effort Choice $h$ by Founder's Effort Cost $c$}
		\label{ch2_fig3}
	\end{center}
\end{figure}

\textit{B. Early stage startups receive more help from venture capitalists.}\\
In this model, venture capitalists' effort choice also depends on the relative effort cost of founders. In Figure \ref{ch2_fig3}, only startups that are more inefficient in management are funded by venture capitalists. Established firms with a complete set of business operations and management structures therefore will not need venture capitalists for financing. On the other hand, early stage startups receive a lot more resources and attention from venture capitalists. Anecdotes often note that early stage startups get hands-on assistance from venture capitalists to set up human resource policies, employee incentives, and other management strategies. \cite{hochberg_accelerating_2016} documents the recent rise in the accelerator program, in which venture capitalists provide mentorship specifically for startups that are just founded.

\textit{C. More startups can be funded when there are more venture capitalists.}\\
The effort cost $v$ for venture capitalists is the shadow cost on the supply constraint. Higher $H$ reduces $v$ and thus increases $h$. \cite{gompers_venture_2020} notes that venture capitalists work about 55 hours a week. There is therefore a capacity constraint on the supply of venture capitalists' effort. When there are more venture capitalists, it will be less costly to spend time on startups. Thus, more startups will be funded and receives more effort on average. \cite{kortum_assessing_2000} analyzes a policy change in 1979 that increases venture capitalist activity, and finds an increase in the number of startups funded and overall higher patent rates.

\section{General Equilibrium}
I embed the partial equilibrium model into a general equilibrium model with free entry of both entrepreneurs and venture capitalists. I characterize the competitive equilibrium and show how venture capitalists' selection and effort choice are affected by entrepreneur types. Then, I analyze the effects of an exogenous decrease in the upfront investment, an exogenous decrease in the interest rate, and an exogenous decrease in the entry cost respectively, and show that the predictions of the model align with empirical observations. 

\subsection{Model Environment}
There is a small open economy that takes risk free rate $r$ as given. It has infinite horizon starting at $t=0$, a representative consumer, a set of penniless risk-neutral entrepreneurs, and a financial sector consisting of risk-neutral bankers and venture capitalists. Entrepreneurs have investment projects that require financing from either bankers or venture capitalists. The projects produce final goods for consumption. The financial sector takes savings from the consumer to fund these projects. The consumer supplies $L$ units of labor.\footnote{I abstract from modeling the consumer, who has a lifetime utility function $\sum_{t=0}^\infty \beta^t c_t^{1-\rho}/(1-\rho)$. }

\subsubsection*{Entrepreneurs}
In the beginning of period $t$, entrepreneurs each receive a draw of an idea with productivity $z \in [\underline{z},\bar{z}]$ and managerial effort cost $c \in [\underline{c},\bar{c}]$. $z$ and $c$ are distributed according to $\Phi(z,c)$. Project requires an upfront investment of $I$. The production function is,
\begin{equation}
	y = z^\chi F^\theta l^\beta, \qquad F = [\gamma h^{\frac{\sigma-1}{\sigma}}
	+ f^{\frac{\sigma-1}{\sigma}} ]^{\frac{\sigma}{\sigma-1}}, \chi + \theta + \beta = 1
\end{equation}
If project is successful, entrepreneur sells the project and gets the present value of the profits in $t+1$. Success probability of the project at the end of period $t$ is $\epsilon$. Subsequent survival rate of the project every period is $s_e$.\footnote{It is commonly observed that success probability during the development period of a project is lower than the survival rate after the project has been successfully developed.}

\subsubsection*{Bankers}
Bankers collect deposits at risk free rate $1+r$. Discount rate, $\delta$, is thus 
\begin{equation}
	\delta = \frac{1}{1+r}    
\end{equation}
The interest rate, $i$, charged by bankers is such that bankers make zero profit, i.e.
\begin{equation}
	1+i = \frac{1}{\delta \epsilon}
\end{equation}
So an entrepreneur who borrows $I$ in period $t$ will repay $I/\delta\epsilon$ in $t+1$.

\subsubsection*{Venture Capitalists}
Each entrepreneur is matched with a particular venture capitalist, and venture capitalist can work with many entrepreneurs. There is no search friction. Upon the match, venture capitalists observe $(z,c)$, and provide $h$ units of expertise to improve the managerial productivity of the project. The aggregate supply of venture capitalists' effort is $H$. The continuation rate of venture capitalists is $s_v$.\footnote{At the end of each period, $1-s_v$ venture capitalists will retire. They stop helping startups and become a pure saver.}

The contract between venture capitalists and entrepreneurs specifies a payment $p$ to venture capitalists in period $t+1$ when project is successful, with venture capitalists' bargaining power being $\alpha$. Neither $h$ or $f$ is observable and cannot be contracted upon. 

\subsubsection*{Entry}
There is free entry with an entry cost $c_e$ and $c_v$ for entrepreneurs and venture capitalists respectively. Entry cost is an increasing function of the number of new entrants, i.e.
\begin{align}
	c_e &= \kappa_e m_e^{\eta_e} \\
	c_v &= \kappa_v m_v^{\eta_v}
\end{align}
where $\kappa_e, \kappa_v$ are positive constants and $m_e,m_v$ are steady state mass of new entrants for entrepreneurs and venture capitalists respectively.

\subsubsection*{Timeline of Events}
In the beginning of time $t$, entrepreneurs draw $(z,c)$ and decide whether or not to undertake the project, and where to seek financing. If entrepreneur goes to the bank, he receives investment $I$, and then puts effort $f$ to work on the project on his own. If entrepreneur goes to the venture capitalist, the two of them first sign a contract specifying $p$, and $I$ is invested. Then, entrepreneur and venture capitalist each chooses his effort level.

At time $t+1$,  project outcome is realized. Successful project produces output according to \ref{eqn_production}, whereas unsuccessful project has zero remaining value. Entrepreneurs repay the bank and venture capitalists.

\subsection{Decision Problems}
Given wage $w$, labor choice $l$ of an ex post successful project is given by the following profit maximization problem,
\begin{equation}
	\Pi \equiv \max_l \{z^\chi F^\theta l^\beta - wl \} = (1-\beta)\Big (\frac{\beta}{w}\Big )^{\frac{\beta}{1-\beta}}z^{\frac{\chi}{1-\beta}} F^{\frac{\theta}{1-\beta}}
\end{equation}

For an entrepreneur who enters at period $t$, the present value of profits from a successful project is
\begin{equation}
	\delta\sum_{t=0}^\infty (\delta s_e)^t\Pi = \frac{\delta}{1-\delta s_e}(1-\beta)\Big (\frac{\beta}{w}\Big )^{\frac{\beta}{1-\beta}}z^{\frac{\chi}{1-\beta}} F^{\frac{\theta}{1-\beta}}
\end{equation}

\subsubsection*{Bank Finance}
Bank-financed entrepreneurs solve the following maximization problem at $t$,
\begin{equation}
	\max_f  \frac{\delta \epsilon}{1-\delta s_e}(1-\beta)\Big (\frac{\beta}{w}\Big )^{\frac{\beta}{1-\beta}}z^{\frac{\chi}{1-\beta}} F^{\frac{\theta}{1-\beta}} - cf - I, \qquad F=f
\end{equation}
Solving for the first order condition for $f$, we have 
\begin{equation}
	f = z \Big ( \frac{\delta \epsilon\theta}{c(1-\delta s_e)}\Big )^{\frac{\theta+\chi}{\chi}} \Big (\frac{\beta}{w}\Big )^{\frac{\beta}{\chi}}
	\label{eqn_fsolo_ge}
\end{equation}
The present value of profits from a bank-financed project and entry threshold is,
\begin{align}
	\label{eqn_vsolo_ge}
	E &\equiv \max \Big\{ z\chi \Big (\frac{\theta}{c} \Big)^{\frac{\theta}{\chi}} \Big (\frac{\delta\epsilon}{1-\delta s_e} \Big)^{1+\frac{\theta}{\chi}}\Big (\frac{\beta}{w} \Big)^{\frac{\beta}{\chi}} - I,0 \Big\} \\
	z_s &\geq \frac{I}{\chi}\Big (\frac{\theta}{c} \Big)^{-\frac{\theta}{\chi}} \Big (\frac{\delta\epsilon}{1-\delta s_e} \Big)^{-1-\frac{\theta}{\chi}}\Big (\frac{\beta}{w} \Big)^{-\frac{\beta}{\chi}}
	\label{eqn_pcsolo_ge}
\end{align}

\subsubsection*{Venture Capital Finance}
The contract between venture capitalists and entrepreneurs is determined by the following Nash Bargaining problem,
\begin{equation}
	\max_{p}[\frac{\delta\epsilon}{1-\delta s_e}(1-\beta)\Big (\frac{\beta}{w}\Big )^{\frac{\beta}{1-\beta}}z^{\frac{\chi}{1-\beta}} F^{\frac{\theta}{1-\beta}} -\epsilon p-E]^{1-\alpha}[\epsilon p - I]^\alpha 
\end{equation}
Solving first order condition for $p$, the surplus for entrepreneur and venture capitalists respectively, is,
\begin{align}
	\label{eqn_valuee_ge}
	S_v &\equiv \epsilon p - I = \alpha(\frac{\delta\epsilon}{1-\delta s_e}(1-\beta)\Big (\frac{\beta}{w}\Big )^{\frac{\beta}{1-\beta}}z^{\frac{\chi}{1-\beta}} F^{\frac{\theta}{1-\beta}} - I - E )    \\
	S_e &= (1-\alpha)(\frac{\delta\epsilon}{1-\delta s_e}(1-\beta)\Big (\frac{\beta}{w}\Big )^{\frac{\beta}{1-\beta}}z^{\frac{\chi}{1-\beta}} F^{\frac{\theta}{1-\beta}} - I - E )  
	\label{eqn_valuevc_ge}
\end{align}
Entrepreneur and venture capitalists maximize surplus respectively given the expectation of the other's effort level. Optimal levels of effort are,
\begin{align}
	f &= \Big (\frac{v}{c}\frac{1-\alpha}{\alpha \gamma}  \Big )^\sigma h
	\label{eqn_fvc_ge} \\
	\label{eqn_hvc_ge}
	h &= z\Big ( \frac{\alpha \delta\epsilon\theta}{v(1-\delta s_e)}\Big )^{\frac{\theta+\chi}{\chi}} \Big (\frac{\beta}{w}\Big )^{\frac{\beta}{\chi}}\gamma^{\frac{\theta \sigma}{\chi(\sigma-1)}}\Big [ 1+\gamma^{-\sigma}\Big(\frac{v}{c}\frac{1-\alpha}{\alpha} \Big)^{\sigma-1}  \Big]^{\frac{\theta}{\chi(\sigma-1)}-1}
\end{align}
Entrepreneur's participation constraint is such that total surplus should be larger than zero,
\begin{equation}
	\label{eqn_pcvc_ge}
	\frac{\delta\epsilon}{1-\delta s_e}(1-\beta)\Big (\frac{\beta}{w}\Big )^{\frac{\beta}{1-\beta}}z^{\frac{\chi}{1-\beta}} F^{\frac{\theta}{1-\beta}} - I \geq E
\end{equation}
Similar to the partial equilibrium model, there is a threshold level $z\geq z_v, c\geq c_v$ above which entrepreneurs will seek venture capitalist financing.

\subsection{Stationary Competitive Equilibrium}
\subsubsection*{Distribution of Entrepreneurs and Venture Capitalists}
To analyze the law of motion for the mass of entrepreneurs, let $M_t$ denote the mass of entrepreneurs at period $t$.
\begin{equation}
	\label{lome}
	M_t = s_e M_{t-1} + \epsilon m_{e,t}
\end{equation}
That is, the mass of entrepreneurs at $t$ consists of surviving entrepreneurs from last period and successful entrants in this period. In steady state, the mass of new entrants is,
\begin{equation}
	m_e = \frac{1-s_e}{\epsilon}M
\end{equation}
In steady state, the entry condition for entrepreneurs is,
\begin{equation}
	\int_{z\geq z_s} E(z,c)d\Phi(z,c) + \int_{z\geq z_vs,c\geq c_v} S_e(z,c)d\Phi(z,c) = c_e = \kappa_e m_e^{\eta_e}
\end{equation}
That is, the expected value of projects (either with bank or with venture capitalists) equals the entry cost.\\
The law of motion for the supply of venture capitalist effort is,
\begin{equation}
	\label{lomvc}
	H_t = s_v H_{t-1} + m_{v,t}
\end{equation}
In steady state
\begin{equation}
	m_v = (1-s_v)H
\end{equation}
Entry condition is,
\begin{equation}
	\int_{z\geq z_v,c\geq c_v} S_v(z,c)d\Phi(z,c) = c_v = \kappa_v m_v^{\eta_v}
\end{equation}

\subsubsection*{Aggregate Quantities}
In aggregate, the demand for labor from successful entrepreneurs add up to the total supply of labor $L$ from the representative consumer. Among the successful entrepreneurs, those with $z\geq z_v,c\geq c_v$ seek venture capitalists financing, while the rest are funded by the bank.
\begin{equation}
	\label{labormarket}
	M \Big(\int_{z\geq z_s} l(z,c)\mathbbm{1}_{\text{Bank}}(z,c) d\Phi(z,c) + \int_{z\geq z_v,c\geq c_v} l(z,c)d\Phi(z,c) \Big) = L
\end{equation}
The demand for venture capitalist effort from new entrants who seek their financing equals the total supply $H$.
\begin{equation}
	m_e \int_{z\geq z_v,c\geq c_v} h(z,c) d\Phi(z,c) = H
	\label{vcmarket}
\end{equation}
Lastly, the demand for funding from new entrants equals the supply of savings $S$ from the representative consumer.
\begin{equation}
	m_e I \Big(\int_{z\geq z_s}\mathbbm{1}_{\text{Bank}}(z,c) d\Phi(z,c) + \int_{z\geq z_v,c\geq c_v} d\Phi(z,c) \Big)  = S
	\label{savingsmarket}
\end{equation}

\subsubsection*{Definition}
An equilibrium consists of prices $\{r,w,v\}$, entry and financing choices of entrepreneurs, effort choice of venture capitalists and entrepreneurs, payment to venture capitalists, labor choice of successful projects, and the distribution over the state space $\mathbb{S}$ given by $\Phi(s), s=(z,c), s\in \mathbb{S}$, such that
\begin{enumerate}
	\item Given wage $w$, successful projects' labor choice satisfies \ref{eqn_labor}.
	\item Given $r,w,v$, entrepreneurs' decision to enter satisfies \ref{eqn_pcsolo_ge}, and the decision to finance with venture capitalists satisfies \ref{eqn_pcvc_ge}. Entrepreneurs' effort choice with bank financing satisfies \ref{eqn_fsolo_ge}, and \ref{eqn_fvc_ge} when with venture capitalists.
	\item Given $r,w,v$, venture capitalists' effort choice satisfies \ref{eqn_hvc_ge}. 
	\item Entrepreneur's surplus when financing with venture capitalists, satisfies \ref{eqn_valuee_ge}, and venture capitalists' surplus satisfies \ref{eqn_valuevc_ge}.
	\item Mass of entrepreneurs evolves according to \ref{lome}.
	\item Mass of venture capitalists evolves according to \ref{lomvc}.
	\item Labor, venture capitalist effort, and savings markets clear in \ref{labormarket}, \ref{vcmarket}, and \ref{savingsmarket}.
\end{enumerate}

\section{Comparative Statics}
\subsection{A Decrease in the Upfront Investment}
To analyze the effects of an exogenous decrease in the upfront investment $I$, I first examine entrepreneur's value under bank financing and his surplus under venture capitalist financing. Notice that,
\begin{equation}
	\frac{\partial z_s}{\partial I} = \frac{1}{\chi}\Big (\frac{\theta}{c} \Big)^{-\frac{\theta}{\chi}} \Big (\frac{\delta\epsilon}{1-\delta s_e} \Big)^{-1-\frac{\theta}{\chi}}\Big (\frac{\beta}{w} \Big)^{-\frac{\beta}{\chi}} > 0
\end{equation}
so entrepreneurs with lower productivity $z$ will now find it more profitable to enter with a decrease in $I$. Moreover, using equation \ref{eqn_valuee_ge} and \ref{eqn_vsolo_ge}, entrepreneur's surplus is,
\begin{multline}
	S_e = \max \Big\{\frac{\alpha\delta\epsilon}{1-\delta s_e}(1-\beta)\Big (\frac{\beta}{w}\Big )^{\frac{\beta}{1-\beta}}z^{\frac{\chi}{1-\beta}} F^{\frac{\theta}{1-\beta}}  -  z\alpha \chi \Big (\frac{\theta}{c} \Big)^{\frac{\theta}{\chi}} \Big (\frac{\delta\epsilon}{1-\delta s_e} \Big)^{1+\frac{\theta}{\chi}}\Big (\frac{\beta}{w} \Big)^{\frac{\beta}{\chi}},\\
	\frac{\alpha\delta\epsilon}{1-\delta s_e}(1-\beta)\Big (\frac{\beta}{w}\Big )^{\frac{\beta}{1-\beta}}z^{\frac{\chi}{1-\beta}} F^{\frac{\theta}{1-\beta}} - I  \Big\}
\end{multline}
where the first part of the surplus is independent of $I$. That is, entrepreneurs who get funded by venture capitalists because it is more profitable than bank financing will not experience a direct impact from a decrease in the upfront cost $I$. On the other hand, entrepreneurs who get funded by venture capitalists but would otherwise not undertake their projects will experience an immediate increase in their surplus following a drop in the cost of starting projects.

Thus, the direct impact of a drop in upfront investment is that more entrepreneurs on the margin will enter to start projects, using either bank or venture capitalist financing.

To analyze the general equilibrium effect of a drop in $I$, first notice that the total mass of entrepreneurs, $M$, increases due to the increased entry of marginal entrepreneurs. This causes equilibrium labor wage, $w$, to increase. Moreover, the increase in new entrants, $m_e$, also makes the supply constraint for venture capitalists' effort tighter, raising the shadow cost $v$. As a result, the amount of effort, $h$, that each venture-funded project receives, decreases, for projects already receiving venture capitalist financing.

To understand the impact on the supply of venture capitalists' effort $H$, I look at the surplus for venture capitalists,
\begin{multline}
	S_v = \max \Big\{\frac{(1-\alpha)\delta\epsilon}{1-\delta s_e}(1-\beta)\Big (\frac{\beta}{w}\Big )^{\frac{\beta}{1-\beta}}z^{\frac{\chi}{1-\beta}} F^{\frac{\theta}{1-\beta}}  -  z(1-\alpha) \chi \Big (\frac{\theta}{c} \Big)^{\frac{\theta}{\chi}} \Big (\frac{\delta\epsilon}{1-\delta s_e} \Big)^{1+\frac{\theta}{\chi}}\Big (\frac{\beta}{w} \Big)^{\frac{\beta}{\chi}},\\
	\frac{(1-\alpha)\delta\epsilon}{1-\delta s_e}(1-\beta)\Big (\frac{\beta}{w}\Big )^{\frac{\beta}{1-\beta}}z^{\frac{\chi}{1-\beta}} F^{\frac{\theta}{1-\beta}} - I  \Big\}
\end{multline} 
where the first part is independent of $I$ and increasing in $h$. That is, surplus collected from entrepreneurs who also would have qualified for bank financing becomes less, since less effort is spent per project with an increase in new entrants. On the other hand, surplus collected from entrepreneurs who would have not entered before now becomes larger, benefiting from a decrease in $I$. The overall effect on $H$ depends on the relative magnitude of the two surpluses. There will be more venture capitalists in equilibrium, provided that the increase in surplus from marginal entrepreneurs outweighs the decrease in surplus from other entrepreneurs.

Lastly, the effect on labor productivity is,
\begin{equation}
	\frac{Y}{L} = \frac{M\int_{z\geq z_s} y(z,c)\mathbbm{1}_{\text{Bank}}(z,c) d\Phi(z,c) + M\int_{z\geq z_v,c\geq c_v} y(z,c)d\Phi(z,c)}{M\int_{z\geq z_s} l(z,c)\mathbbm{1}_{\text{Bank}}(z,c) d\Phi(z,c) + M\int_{z\geq z_v,c\geq c_v} l(z,c)d\Phi(z,c)} = \frac{w}{\beta}
\end{equation}
since for every operating business, the first order condition with respect to $l$ gives,
\begin{equation}
	w = \beta z^\chi F^\theta l^{\beta-1} = \frac{y}{l}
\end{equation}
Thus labor productivity is proportional to $w$ and hence increases.

I summarize the results as follows.
\begin{proposition}
	A decrease in the upfront cost of investment $I$ leads to
	\begin{enumerate}
		\item a reduction in the entry threshold $z_s,z_v$.
		\item a decrease in effort received from venture capitalists $h$.
		\item an increase in the total mass of entrepreneurs $M$.
		\item an increase in labor productivity $\frac{Y}{L}$.
	\end{enumerate}
\end{proposition}

\begin{figure}[H]
	\begin{center}
		\includegraphics[width=\textwidth]{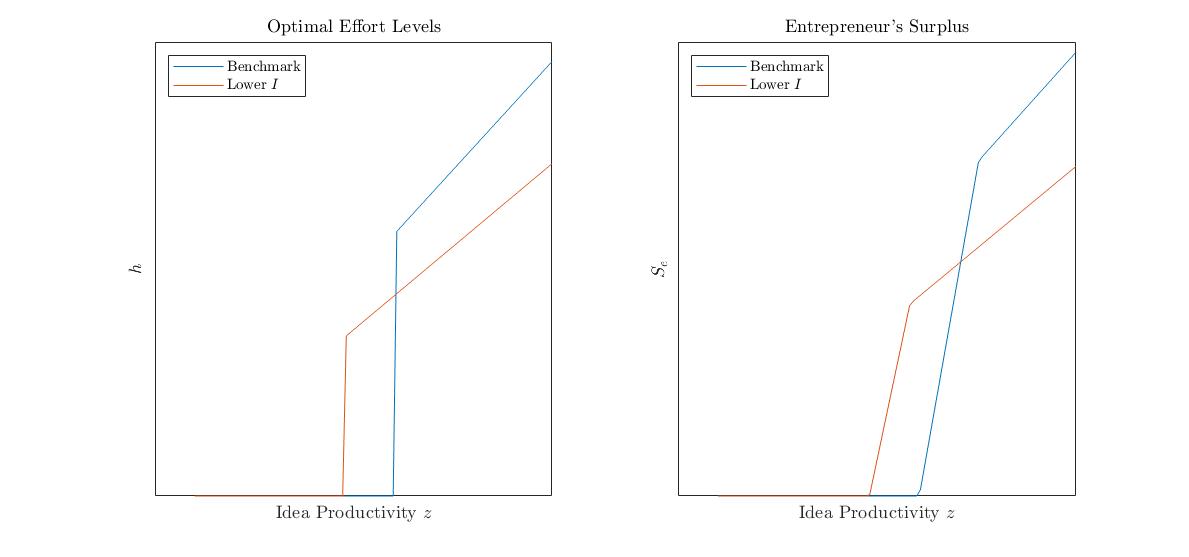}
		\caption{Effects of decrease in $I$ - effort levels and entrepreneur's surplus.}
		\label{ch2_fig4}
	\end{center}
\end{figure}

Figure \ref{ch2_fig4} illustrates the effects of a decrease in the upfront investment $I$ using a numerical example. Details can be found in Appendix \ref{ch2_appendix}. Equilibrium values before the change are indicated with blue solid lines while those after the change are represented by red solid lines. For a given founder's cost $c$, projects with lower $z$ get funded and receive more surplus than before. However, projects already funded before the change now receive less effort and surplus.

\begin{figure}[H]
	\begin{center}
		\includegraphics[width=0.6\textwidth]{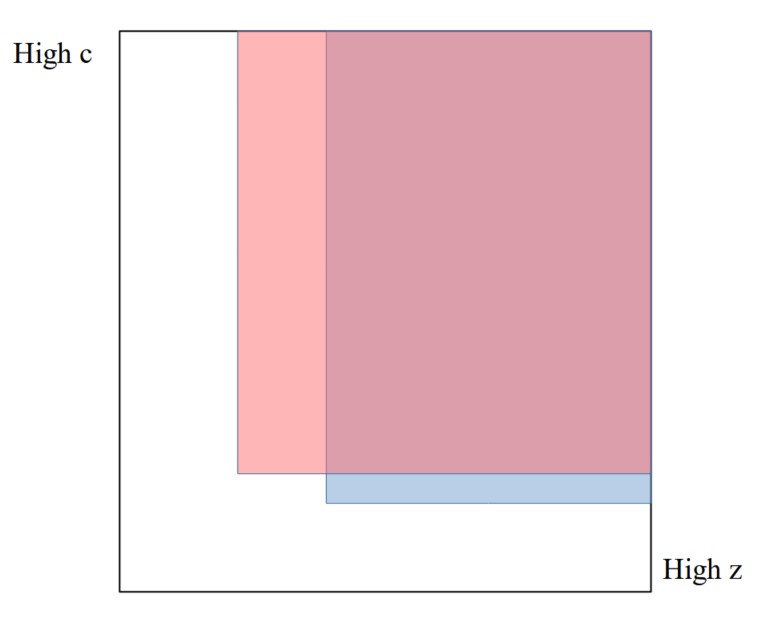}
		\caption{Effects of decrease in $I$ - venture-funded region}
		\label{ch2_fig5}
	\end{center}
\end{figure}

Figure \ref{ch2_fig5} shows the venture-funded region before and after the decrease in $I$. The region funded before the change is indicated with blue-shaded area, while the region funded after is the red-shaded area. Entrepreneurs with lower $z$ benefit from the decrease in the upfront investment. On the other hand, entrepreneurs with lower $c$ now opt out of venture capitalists' service, as the cost of venture capitalists' effort becomes higher due to an increase in demand for their effort.

In \cite{ewens_cost_2018}, the authors analyze changes in investment approach by venture capitalists following the advent of Amazon's Web Services (AWS) in early 2006. Cloud computing lowers the initial cost of starting businesses for firms in online services. Using a difference-in-difference approach, the authors find that startups affected by cloud computing require a lower initial funding size ($I$). Moreover, venture capitalists have made more investments, while providing less governance ($h$) to their portfolio companies. The authors further note that marginal ventures with lower expected payoffs comprise of most of the new investments made by venture capitalists. Their observations align with the model predictions on the effects of a decrease in $I$.

\subsection{A Decrease in Interest Rates}
A decrease in interest rate, $r$, increases the discount rate $\delta$. The direct effect of an increase in $\delta$ is,
\begin{equation}
	\frac{\partial E}{\partial \delta }> 0, \qquad \frac{\partial S_e}{\partial \delta }> 0, \qquad \frac{\partial S_v}{\partial \delta }> 0, \qquad \frac{\partial h}{\partial \delta } > 0
\end{equation}
That is, higher present value of projects leads to higher expected payoff when financing with the bank, higher surplus for both entrepreneurs and venture capitalists, and motivates venture capitalists to provide more effort. 

In equilibrium, higher surplus induces more entrepreneurs and venture capitalists to enter, raising the total mass $M$ and $H$. This in turn leads to higher demand for labor and venture capitalist effort, raising equilibrium prices $v$ and $w$. Higher $v,w$ reduces effort $h$ and expected payoff and surplus $S_e, S_v, E$. Thus, the overall effect on effort levels $h$ is less clear. Labor productivity increases as wage increases.
\begin{proposition}
	A decrease in interest rate $r$ leads to
	\begin{enumerate}
		\item a reduction in the entry threshold $z_s,z_v$.
		\item an increase in the total mass of entrepreneurs $M$.
		\item an increase in the total supply of venture capitalists effort $H$.
		\item an increase in the labor productivity, $\frac{Y}{L}$.
	\end{enumerate}
\end{proposition}

\begin{figure}[H]
	\begin{center}
		\includegraphics[width=\textwidth]{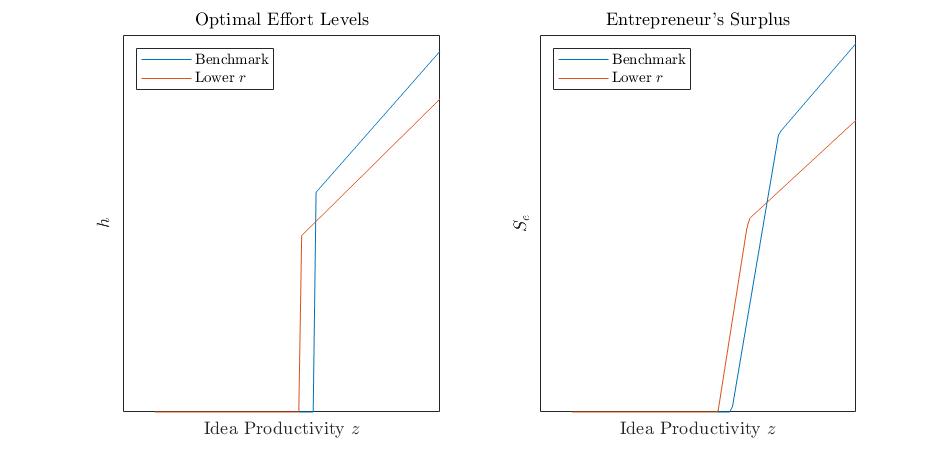}
		\caption{Effects of decrease in $r$ - effort levels and entrepreneur's surplus.}
		\label{ch2_fig8}
	\end{center}
\end{figure}

Figure \ref{ch2_fig8} shows the optimal effort level and entrepreneur's surplus before and after the decrease in interest rate $r$. Given the parameters, the negative impact on $h$ from an increased $w,v$ significantly outweighs the positive impact from a decrease in $r$. As a result, while the newly funded projects get attended by venture capitalists compared to not funded at all, the projects that would have been funded before the decrease in $r$ now receives less attention. As entrepreneur's surplus is an increasing function of $h$, the effect on entrepreneur's surplus on the right panel exhibits the same pattern as the left panel.

A recent Economist article\footnote{https://www.economist.com/finance-and-economics/2021/11/23/the-bright-new-age-of-venture-capital/21806438} notes that the fall in interest rates has prompted more investors to enter into venture capital financing, investing more funds into riskier businesses. This is in line with model predictions, that lower interest rates encourages the entry of marginal entrepreneurs and also venture capitalists.

\subsection{A Decrease in Venture Capitalists' Entry Cost}
When regulatory changes lower venture capitalists' entry cost, $\kappa_v$, the immediate effect is to raise the amount of venture capitalists' effort, $H$. From the entry condition, we have,
\begin{equation}
	H = \frac{1}{1-s_v}\Big (\frac{\int_{z\geq z_v,c\geq c_v} S_v(z,c)d\Phi(z,c)}{ \kappa_v} \Big)^{\frac{1}{\eta_v}} 
\end{equation}
so that $\frac{\partial H}{\partial \kappa_v}<0$. Moreover, from the market clearing condition,
\begin{equation}
	m_e \int_{z\geq z_v,c\geq c_v} h(z,c) d\Phi(z,c) = H
\end{equation}
so that a higher supply of venture capitalists' effort lowers the shadow cost $v$. 

The shadow cost $v$ affects venture-funded projects in two ways. On the extensive margin, founders who have lower management cost $c$ and used to find venture capitalists' effort more expensive will now find it more profitable to finance with venture capitalists than the bank given the lower shadow cost $v$. On the intensive margin, each venture-funded project now gets more effort, since,
\begin{equation}
	h = z\Big ( \frac{\alpha \delta\epsilon\theta}{v(1-\delta s_e)}\Big )^{\frac{\theta+\chi}{\chi}} \Big (\frac{\beta}{w}\Big )^{\frac{\beta}{\chi}}\gamma^{\frac{\theta \sigma}{\chi(\sigma-1)}}\Big [ 1+\gamma^{-\sigma}\Big(\frac{v}{c}\frac{1-\alpha}{\alpha} \Big)^{\sigma-1}  \Big]^{\frac{\theta}{\chi(\sigma-1)}-1}, \qquad \frac{\partial h}{\partial v} < 0
\end{equation}
As each venture-funded project gets higher effort $h$, managerial productivity $F$ becomes higher. The marginal product of labor thus increases, raising the demand for labor. Wage $w$ therefore increases, and so is labor productivity.

On the other hand, the effect on the total mass of entrepreneurs, $M$, is unclear. Although venture-funded entrepreneurs get a higher surplus due to higher effort, bank-funded entrepreneurs may get lower payoffs due to higher wages. Summarizing the results, we have
\begin{proposition}
	A decrease in venture capitalists' entry cost $\kappa_v$ leads to
	\begin{enumerate}
		\item a reduction in the entry threshold $c_v$ for venture financing.
		\item an increase in effort $h$ spent on each project.
		\item an increase in the total supply of venture capitalists effort $H$.
		\item an increase in labor productivity, $\frac{Y}{L}$.
	\end{enumerate}
\end{proposition}

\begin{figure}[H]
	\begin{center}
		\includegraphics[width=\textwidth]{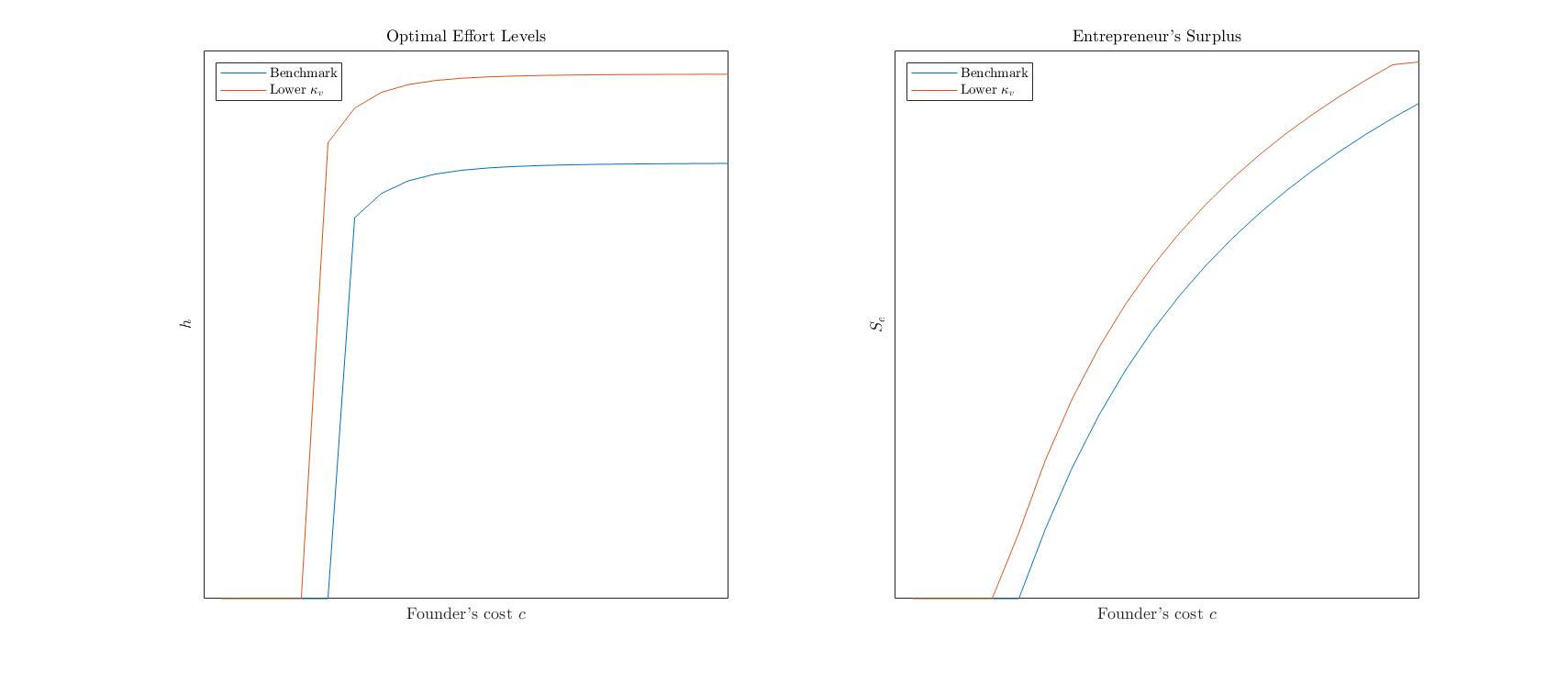}
		\caption{Effects of decrease in $\kappa_v$ - optimal effort and entrepreneur's surplus}
		\label{ch2_fig6}
	\end{center}
\end{figure}

\begin{figure}[H]
	\begin{center}
		\includegraphics[width=0.6\textwidth]{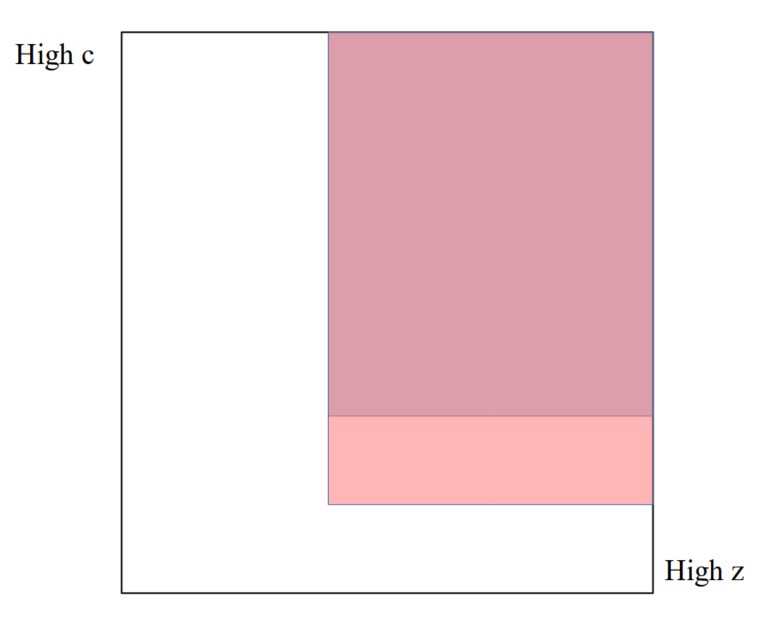}
		\caption{Effects of decrease in $\kappa_v$ - venture-funded region}
		\label{ch2_fig7}
	\end{center}
\end{figure}

Figures \ref{ch2_fig6} and \ref{ch2_fig7} illustrates the effect of a lower entry cost $\kappa_v$ from the numerical example. Figure \ref{ch2_fig6} shows how venture capitalists' effort provision and entrepreneur's surplus vary by founder's effort cost $c$, for a given feasible productivity level $z$. Each project now gets more effort and higher surplus for the entrepreneur. Figure \ref{ch2_fig7} further shows that the entry threshold $c_v$ is now lower following a decrease in the entry cost.

The predictions of the model are in line with the recent findings in \cite{ewens_deregulation_2020}. The authors analyze the impact of a deregulation event, the National Securities Market Improvement Act (NSMIA), on venture capitalists' funding behavior. NSMIA has made it easier for startups to raise private capital from out-of-state investors in a few ways, such as exempting private firms from complying with the blue sky laws of every new state where they issue securities. The removal of restrictions reduces the entry cost especially for out-of-state investors. The authors find an increase in the funding amount to late-stage startups. Moreover, they document an increase in scale in startups, as seen from higher employment and sales. In my model, late-stage startups resemble low $c$ type entrepreneurs\footnote{Late-stage startups often have already developed a mature system of business operations, hence the cost to raise management productivity is lower.}, and therefore benefit from a lower entry cost. The increase in employment is also reflected in my model, as higher effort provision by venture capitalists raise the management productivity of projects and subsequently labor demand becomes higher.

\section{Conclusion}
This paper writes a general equilibrium model that focuses on venture capitalists' effort supply and its impact. The key assumption is that venture capitalists' effort increases firm's managerial productivity, and its supply is determined using a free entry condition. Model predictions are in line with several empirical observations, so are the comparative statics. My model shows that although venture capitalists have positive effects on labor productivity, they face a supply constraint that limits their ability to assist more firms. This study has two implications. First, it shows the importance of understanding the time use of venture capitalists, and urges future data collection effort to better understand how venture capitalists optimize their effort among different types of firms and investment activities. Second, it shows that under current low interest rate environment, policymakers should focus more on venture capitalists' time constraint, rather than the inflow of funds, as the former is more likely to be the bottleneck to the growth of entrepreneurship.

\section*{Appendix}\label{ch2_appendix}
\subsection{Merging NETS and VentureXpert}
Before carrying out the merging procedure, I apply standardization to firm identifiers in VentureXpert and NETS. All firm names are changed to uppercase, and unwanted spaces are removed. The suffixes for firm names are standardized to an abbreviated form, for example, "Limited" is changed to "LTD", and "Corporation" is changed to "CO", and so on. The suffixes for firm addresses are similarly standardized, for example, "Drive" is changed to "DR". Lastly, I create a list of unique firm identifiers for each dataset, by dropping duplicates of name, state, city, zipcode, and SIC combinations. 

The merging procedure is carried out in two steps. The first step is direct merge using identifiers, in nine passes. The first pass uses all identifiers, which are name, state, city, zipcode, address, and SIC. The second pass leaves out address. The third pass uses name, state, city, and SIC. The fourth pass uses name, state, zipcode, and SIC. The fifth to seventh pass use name and SIC, plus one of the identifiers from city, zipcode, and state. The eighth pass uses name and SIC. And the last pass matches on name.

The second step is fuzzy names matching using the Python package \textit{fuzzymatcher}. The algorithm generates a probabilistic match score in between 0 and 1 for each pair of names. To keep only one match, I pick the matched pair with the maximum score if there are multiple matched pairs. Moreover, I set the minimum score to be 0.5. 

A filtering procedure is applied after the merging procedure, to verify the matched NETS establishment is indeed venture-funded. First, I check the SIC codes for the matched pairs to make sure they are identical for the first two digits. Second, I drop matched pairs if the year that the establishment first appears in NETS and the year that the firm is first founded in VentureXpert is more than five years apart. Lastly, I drop matched pairs if the last year that the establishment appears in NETS is more than five years earlier than the last year that the firm receives funding in VentureXpert. The percentage of merged firms by merge type is presented in Table \ref{ch2_appendix_table}

\begin{table}[H]
	\centering
	\caption{Composition of Merged Names in NETS and VentureXpert}
	\begin{tabular}{lccc}
		\hline
		Merge Type & Frequency & Percent & Cumulative \\
		name, state, city, zipcode, address, SIC & 483 & 2.70 & 2.70 \\
		name, state, city, zipcode, SIC & 1323 & 7.38 & 10.08 \\
		name, state, city, SIC & 343 & 1.91 & 11.99\\
		name, state, zipcode, SIC & 44 & 0.25 & 12.24 \\
		name, state, SIC & 688 & 3.84 & 16.08 \\
		name, SIC & 742 & 4.14 & 20.22\\
		fuzzy name & 1133 & 6.32 & 26.54 \\
		name & 13165 & 73.46 & 100.00\\
		\hline
		Total & 17921 & 100.00 & 100.00\\
		\hline
		
	\end{tabular}
	
	\label{ch2_appendix_table}
\end{table}
17921 venture-funded firms are matched with a NETS establishment. This corresponds to a 50\% match rate, given the total number of unique venture-funded firms is 35675.

\subsubsection{Limitation of NETS}
NETs is a private source of US business microdata. Previous studies (\cite{barnatchez_assessment_2017},\cite{crane_business_2019}) have compared NETS to official sources of business data and reached the following conclusions. First, NETS uses a different definition for business, and thus have a different establishment count compared to Census. This will underestimate the percentage of venture-funded establishments, if majority of them are too young or too small to be counted in the NETS definition. \cite{puri_life_2012} counts 0.11\% of newly founded venture firms in NETS, whereas NETS only counts 0.01\%.

Second, business dynamics in NETS is "muted". NETS misses a lot of dispersion in employment growth for young firms, mostly due to the large amount of missing employment data in firms' early years. Following the literature's guide, I refrain from drawing conclusions using growth data. Moreover, I also do not use sales data as it is entirely imputed from employment and has little dispersion.

\subsection{Proofs}
\subsubsection{Proof for Proposition \ref{prop1}}
In equation \ref{eqn_hvc}, differentiating $h$ with respect to $z,c,v$ respectively, we have
\begin{multline}
	\frac{\partial h}{\partial z} = \Big ( \frac{\alpha \epsilon\theta}{v(1+r)}\Big )^{\frac{\theta+\chi}{\chi}} \Big (\frac{\beta}{w}\Big )^{\frac{\beta}{\chi}}\gamma^{\frac{\theta \sigma}{\chi(\sigma-1)}}\Big [ 1+\gamma^{-\sigma}\Big(\frac{v}{c}\frac{1-\alpha}{\alpha} \Big)^{\sigma-1}  \Big]^{\frac{\theta}{\chi(\sigma-1)}-1} > 0
\end{multline}
\begin{multline}
	\frac{\partial h}{\partial c} = z\Big ( \frac{\alpha \epsilon\theta}{v(1+r)}\Big )^{\frac{\theta+\chi}{\chi}} \Big (\frac{\beta}{w}\Big )^{\frac{\beta}{\chi}}\gamma^{\frac{\theta \sigma}{\chi(\sigma-1)}}\Big [ 1+\gamma^{-\sigma}\Big(\frac{v}{c}\frac{1-\alpha}{\alpha} \Big)^{\sigma-1}  \Big]^{\frac{\theta}{\chi(\sigma-1)}-2}\\
	\Big(\sigma-1-\frac{\theta}{\chi} \Big)\gamma^{-\sigma}\Big(\frac{v(1-\alpha)}{\alpha} \Big)^{\sigma-1}c^{-\sigma} > 0, \qquad \text{if} \quad \frac{\theta}{\chi}<\sigma-1 
\end{multline}
\begin{multline}
	\frac{\partial h}{\partial v} = -\Big(1+\frac{\theta}{\chi}\Big)z\Big ( \frac{\alpha \epsilon\theta}{(1+r)}\Big )^{\frac{\theta}{\chi}} \Big (\frac{\beta}{w}\Big )^{\frac{\beta}{\chi}}\gamma^{\frac{\theta \sigma}{\chi(\sigma-1)}}\Big [ 1+\gamma^{-\sigma}\Big(\frac{v}{c}\frac{1-\alpha}{\alpha} \Big)^{\sigma-1}  \Big]^{\frac{\theta}{\chi(\sigma-1)}-1} + \\
	z\Big ( \frac{\alpha \epsilon\theta}{v(1+r)}\Big )^{\frac{\theta+\chi}{\chi}} \Big (\frac{\beta}{w}\Big )^{\frac{\beta}{\chi}}\gamma^{\frac{\theta \sigma}{\chi(\sigma-1)}}\Big [ 1+\gamma^{-\sigma}\Big(\frac{v}{c}\frac{1-\alpha}{\alpha} \Big)^{\sigma-1}  \Big]^{\frac{\theta}{\chi(\sigma-1)}-2}\\
	\gamma^{-\sigma}\Big(\frac{(1-\alpha)}{c\alpha} \Big)^{\sigma-1}\Big(\frac{\theta}{\chi}+1-\sigma \Big)v^{\sigma-2} < 0, \qquad \text{if} \quad \frac{\theta}{\chi}<\sigma-1 
\end{multline}
where the condition $\frac{\theta}{\chi}<\sigma-1$ is satisfied given parameter estimates of $\theta,\chi$, and $\sigma$ in \cite{bloom_management_2016}.

\subsubsection{Feasibility Region}
From equation \ref{eqn_pcsolo}, we have
\begin{equation}
	\frac{\partial z_s}{\partial c} = \frac{\theta}{\chi}\frac{I}{\chi}\theta^{-\frac{\theta}{\chi}} c^{\frac{\theta}{\chi}-1} \Big (\frac{\epsilon}{1+r} \Big)^{-1-\frac{\theta}{\chi}}\Big (\frac{\beta}{w} \Big)^{-\frac{\beta}{\chi}} > 0
\end{equation}
Thus, the higher $c$ is, the higher the threshold $z_s$ for the project to be feasible under bank financing. This gives the blue dashed line in Figure \ref{ch2_fig1}.

Using equation \ref{eqn_pcvc} and equation \ref{eqn_pcsolo}, we have,
\begin{equation}
	\frac{\epsilon}{1+r}(1-\beta)\Big (\frac{\beta}{w}\Big )^{\frac{\beta}{1-\beta}}z^{\frac{\chi}{1-\beta}} F^{\frac{\theta}{1-\beta}} - I \geq \max \Big \{ z\chi \Big (\frac{\theta}{c} \Big)^{\frac{\theta}{\chi}} \Big (\frac{\epsilon}{1+r} \Big)^{1+\frac{\theta}{\chi}}\Big (\frac{\beta}{w} \Big)^{\frac{\beta}{\chi}} - I, 0 \}
\end{equation}
When the project is feasible under bank financing, that is, $z\chi \Big (\frac{\theta}{c} \Big)^{\frac{\theta}{\chi}} \Big (\frac{\epsilon}{1+r} \Big)^{1+\frac{\theta}{\chi}}\Big (\frac{\beta}{w} \Big)^{\frac{\beta}{\chi}} - I > 0$, we have,
\begin{equation}
	\frac{\epsilon}{1+r}(1-\beta)\Big (\frac{\beta}{w}\Big )^{\frac{\beta}{1-\beta}}z^{\frac{\chi}{1-\beta}} F^{\frac{\theta}{1-\beta}} - I \geq  z\chi \Big (\frac{\theta}{c} \Big)^{\frac{\theta}{\chi}} \Big (\frac{\epsilon}{1+r} \Big)^{1+\frac{\theta}{\chi}}\Big (\frac{\beta}{w} \Big)^{\frac{\beta}{\chi}} - I
\end{equation}
Substitute in the expression for $F$, the inequality becomes,
\begin{align}
	&z\frac{\epsilon}{1+r}(1-\beta)\Big (\frac{\beta}{w}\Big )^{\frac{\beta}{1-\beta}}\Big \{(\theta+\chi)\Big(\frac{\alpha \theta}{v} \Big)^{\frac{\theta}{\chi}}\gamma^{\frac{\sigma\theta}{\chi(\sigma-1)}} \Big [ 1+\gamma^{-\sigma}\Big(\frac{v}{c}\frac{1-\alpha}{\alpha} \Big)^{\sigma-1}  \Big]^{\frac{\theta}{\chi(\sigma-1)}} - \chi\Big (\frac{\theta}{c} \Big)^{\frac{\theta}{\chi}}  \Big\} \geq 0\\
	&\frac{\theta+\chi}{\chi}c^{\frac{\theta}{\chi}}\Big(\frac{\alpha}{v} \Big)^{\frac{\theta}{\chi}}\gamma^{\frac{\sigma\theta}{\chi(\sigma-1)}} \Big [ 1+\gamma^{-\sigma}\Big(\frac{v}{c}\frac{1-\alpha}{\alpha} \Big)^{\sigma-1}  \Big]^{\frac{\theta}{\chi(\sigma-1)}} \geq 1
\end{align}
Equation (B.8) says that among projects that are feasible under bank financing, there exists a threshold $c_v$ that separates out projects that are more profitable under venture capitalist financing. Further differentiate the left hand side (LHS) of equation (B.8) with respect to $c$ and leave out constant terms, we have,
\begin{align}
	\frac{\partial \text{LHS}}{\partial c} = \Big [ 1+\gamma^{-\sigma}\Big(\frac{v}{c}\frac{1-\alpha}{\alpha} \Big)^{\sigma-1}  \Big]- \gamma^{-\sigma}\Big(\frac{v(1-\alpha)}{\alpha} \Big)^{\sigma-1}c^{1-\sigma} = 1 > 0
\end{align}
Thus, the left hand side is strictly increasing in $c$. For $c\geq c_v$, projects that are already feasible under bank financing will be more profitable under venture capitalist financing. 

When the project is not feasible under bank financing, but venture capitalists are more efficient, this makes,
\begin{equation}
	\frac{\epsilon}{1+r}(1-\beta)\Big (\frac{\beta}{w}\Big )^{\frac{\beta}{1-\beta}}z^{\frac{\chi}{1-\beta}} F^{\frac{\theta}{1-\beta}} - I \geq 0
\end{equation}
This inequality gives combinations of $(z,c)$ that is feasible under venture capitalist financing but not under bank financing.

\subsection{Numerical Example}
To illustrate the effect of changes in certain parameters, I solve the general equilibrium model. Although some model parameters have to come from a priori assumptions, the following parameters in Table \ref{geparams} can be empirically determined. 

\begin{table}[H]
	\centering
	\caption{Main Parameters}
	\label{geparams}
	\begin{tabular}{lcc}
		\hline
		Parameter &       Value&        Source\\
		\hline
		$\sigma$     &5 &   \cite{bloom_management_2016}\\
		$\theta$  &   0.1 &     \cite{bloom_management_2016}\\
		$\beta$ & 0.6 &\cite{bloom_management_2016} \\
		$\epsilon$ & 0.57 & VentureXpert
		1985-2015\footnote{This includes mergers and acquisitions. IPO rate is 17\%.} \\
		$s_e$  &  0.96 &  \cite{akcigit_synergizing_2019}\\
		$s_v$ & 0.8 & VentureXpert 1985-2015 \\
		\hline
	\end{tabular}
\end{table}

\subsubsection{Computation Algorithm}
\begin{enumerate}
	\item Guess the total mass of entrepreneurs and venture capitalists, $M,H$. 
	\item Guess wage $w$.
	\item Given $H$, solve for $h(z,c)$ by guessing shadow cost $v$ so that the demand for venture capitalists effort equals supply.
	\item Check if labor market clears, if not, go back to Step 2 to update the guess for $w$.
	\item Check if entrepreneur and venture capitalists' expected value of entry equal to the entry cost, if not, go back to Step 1 to update the guess for $M,H$.
\end{enumerate}

The equilibrium values of the numerical examples are listed below in Table \ref{ch2_table2}.

\begin{table}[H]
	\centering
	\caption{Steady State Values}
	\begin{tabular}{lcccc}
		\hline
		& Benchmark & Lower $I$ &  Lower $r$ & Lower $\kappa_v$ \\
		\hline
		$w$ & 1.14 & 1.26 & 1.26 & 1.15 \\
		$v$ & 1.27 & 1.38 & 1.54 & 1.13 \\
		$H$  & 1.36 & 1.41  & 1.37 & 1.63 \\
		$M$ & 10.54  &  12.29  & 12.44  & 10.58\\
		$\frac{Y}{L}$ & 1.90 & 2.11 & 2.10 & 1.92 \\
		\hline
	\end{tabular}
	\label{ch2_table2}
\end{table}

\begin{singlespace}

	\bibliographystyle{econometrica}
	\bibliography{thesis_vc,masterbib}
\end{singlespace}

\end{document}